# Techno-economic Analysis of Light Isotope-enriched Elements for Lightweighting Applications

Wenbo Bao[1], Zhihao Yang[1], Taeyoung Wang[1], Joseph F. Wild[1], Yuan Yang[1]*

[1]Department of Applied Physics and Applied Mathematics, Columbia University, New York, NY 10027, USA. *Corresponding author. Email: yy2664@columbia.edu (Yuan Yang).

## Abstract

Lightweighting is critical to mass-sensitive applications such as aircraft and space transportation. Conventional lightweight strategies often rely on new designs of materials and structures. An alternative approach is to enrich the lightest stable isotopes in an element to reduce the element's atomic mass while having little effect on structural and chemical properties. However, the economic feasibility of this concept remains unclear. Here we present a techno-economic analysis of light isotope-enriched elements for lightweighting applications by estimating isotope enrichment cost and the economic gain from mass reduction. The enrichment cost is scaled from established large-scale processes. 12 common aerospace-relevant elements are considered, including Li, B, C, Mg, Cl, Ti, Ni, Fe, Cu, Zn, Mo, and Sn. We find that 9 elements, especially Li, B, Zn, Ni, Mo, and Sn show potentially attractive economic benefit at moderate enrichment levels, whereas C, Mg, and Fe provide little or no benefit. With the optimized enrichment levels, an Airbus A380 is expected to save approximately \$700 K over a 30-year operational lifetime, a SpaceX Falcon could save \$516K, and a SpaceX starship is expected to save \$2.37 million over its whole lifetime. While the exact enrichment cost needs to be further investigated, these results provide an initial screening of promising candidate elements and highlight isotopic mass reduction as a potential drop-in lightweighting strategy.

## 1. Introduction

Lightweighting is critical to mass-sensitive applications, such as aircraft and space transportation. Conventional lightweighting materials include low-density alloys[1, 2], carbon fiber-based composites[3, 4], and high-performance coatings[5], which often require material substitution and redesign to meet harsh requirements in multiple dimensions (e.g., extreme temperatures, chemical stability, structural robustness). An alternative and rarely explored approach is to use materials that are enriched in their lightest stable isotope to decrease atomic mass with little impact on structural and chemical properties. For example, Ti has five isotopes, $^{46}$Ti (8.3%), $^{47}$Ti (7.4%), $^{48}$Ti (73.7%), $^{49}$Ti (5.4%), $^{50}$Ti (5.2%). Using pure $^{46}$Ti can reduce the density of Ti by 4.0%, which represents a mass reduction of 1,104 kg for an A380 airplane[6, 7]. We only find one patent proposing this concept in literature[8]. However, there is no study to quantitatively analyze the feasibility and economic benefits of this concept.

This concept is attractive as it does not introduce any new materials and is a drop-in solution. To evaluate the feasibility of this concept, it is critical to first analyze the economic gain, since enriched isotopes are expensive, ranging from ~\$1-10/g for those enriched by chemical or distilling methods (e.g., D, $^{6}Li$)[9, 10] to ~$100/g for those enriched by gas centrifugation (e.g., $^{235}U$)[11, 12] and $$10^4$-$10^5$/g for those enriched by electromagnetic separation (e.g., $^{176}Yb$)[13, 14]. On the other side, in aircraft and space transportation, the cost saving per mass is in the order of ~$10,000/kg in the whole life.

In this study, we present the first analysis on how enriched lightweight isotopes can benefit aerospace economy by estimating the cost of isotope enrichment and the fuel saving as a result of weight reduction. 12 common elements used in aerospace vehicle are considered including Li, B, C, Mg, Cl, Ti, Ni, Fe, Cu, Zn, Mo, and Sn. We found that all elements except for C, Mg, and Fe show potentially attractive economic benefit if moderately enriched, such as $22/kg for Ni with $^{58}Ni$ enriched from 68.9% to 75.8%, and $55/kg for Zn with $^{64}Zn$ enriched from 50.2% to 66.0%. It is estimated that a single A380 aircraft could save approximately $764 K over a 30-year operational lifetime. Moreover, applying such light isotope-enriched elements could also result in extra $516 K and $2.37 M profit for SpaceX Falcon 9 and SpaceX Starship, respectively. Besides, such light-isotope-enriched elements could be available from the tail of existing enrichment systems, such as Zn, Mo and Ni[15-17]. On the other hand, it should be cautioned that the enrichment cost is based on a couple of ideal assumptions, such as ton-scale production, and simple scaling from matured isotope separation technologies, which may not be precise. We consider our analysis as a semi-quantitative initial estimation of the potential of enriched isotopes for lightweighting applications.

## 2. Modeling

## 2.1 The analysis framework

The atomic percentage of different isotopes in an element $E$ is denoted as $\vec{x} = [x_1, x_2, \ldots x_k]$, where 1 to $k$ represent isotopes from the lowest to the highest atomic mass, and $\vec{x}_{nat}$ correspond to the natural abundance. Then $B(\vec{x})$, the benefit of using the light-isotope-enriched element $E$ with a composition of $\vec{x}$, can be expressed as

$$B(\vec{x}) = G(1 - r(\vec{x})) - C(\vec{x}) \quad (1)$$

where $G$ is the gain of saving a mass of 1 kg in the entire lifetime, $C$ is the cost to enrich lightweight isotopes so the composition of $E$ changes from $\vec{x}_{nat}$ to $\vec{x}$, and $r$ is the molar mass ratio of $E$ with a composition of $\vec{x}$ to that with the natural composition $\vec{x}_{nat}$. $r$ can be simply expressed as

$$r = \frac{\sum m_i x_i}{\sum m_i x_{nat,i}} \quad (2)$$

Where $m$ (g/mol) is the molar mass of an isotope, and $i$ is over all isotopes in this element. $G$ can be estimated as $M \times P \times yr$, where $M$ is the mass of fuel saved per kg of mass saved per year, $P$ is the fuel price ($/kg), and $yr$ is the number of operational years. In this study, an average $G$ of $13.2K/kg is adopted, assuming a 30-year lifetime with two trips per day[18, 19] (see S.I. section S1).

Isotope separation typically consists of multiple stages in series in a cascade, as shown in Fig. 1, so $C(\vec{x})$ is the sum of cost in each stage. In the $n$ th stage, $C_n(\vec{x})$ can be considered as the product of the separative work unit (SWU) to enrich $E$ to the composition $\vec{x}$ and the cost per SWU ($/SWU, or $C_{SWU}$)[20].

$$C = \sum_n SWU(n) \times C_{SWU}(n) \quad (3)$$

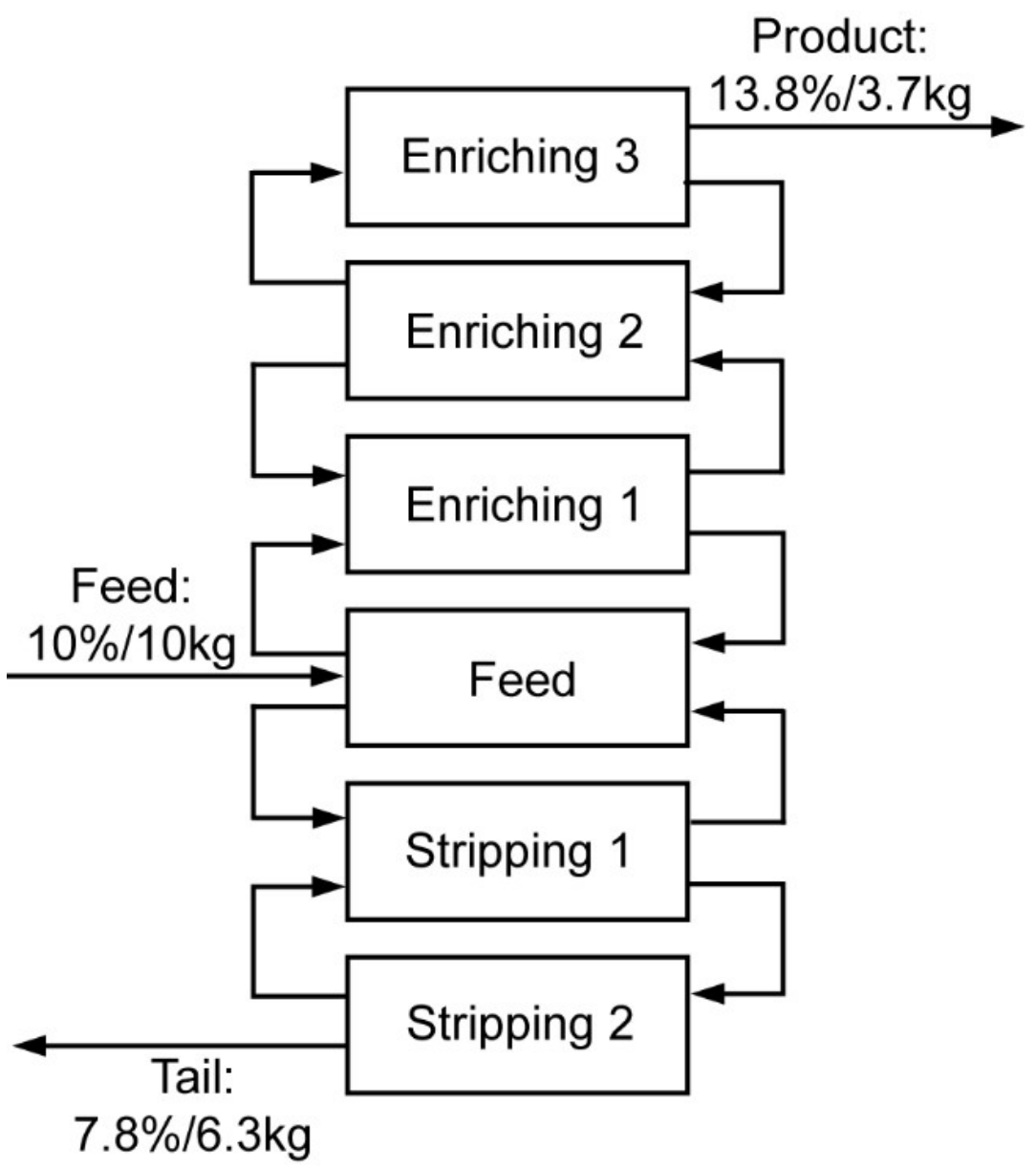


Figure 1. Schematic diagram of a cascade for an artificial isotope separation, with the natural abundance of lightest isotope at 10% and a separation factor $\alpha$ of 1.2. It consists of three enriching stages, two stripping stages, and one feeding stage. Upward arrows: product stream to the upper stage; downward arrows: tails to the lower stage. If the feed mass ($M_F$) is 10 kg with an abundance of the light isotope ($x_{1,F}$) of 10%, the product mass ($M_P$) is 3.7 kg with an abundance of the lightest isotope ($x_{1,P}$) of 13.8%, and the tail mass ($M_T$) is 6.3 kg with an abundance of the lightest isotope ($x_{1,T}$) of 7.8%.

**2.2 SWU per stage and cost per SWU**

In each stage, the separator separates the feed with a mass of $M_F$ and isotopic abundance of $\vec{x}_F$ to the product (lightest isotope enriched) with $M_P$ and $\vec{x}_P$, and the tail (lightest isotope depleted) with the mass of $M_T$ and $\vec{x}_T$ (Fig. 1). Since we care about the lightest isotope, then

the SWU of a stage can be calculated with a focus of the lightest isotope, whose abundance are $x_{1,F}$, $x_{1,P}$, and $x_{1,T}$ in feed, product, and tail, respectively.

$$SWU = M_P V(x_{1,P}) + M_T V(x_{1,T}) - M_F V(x_{1,F}) \quad (4)$$

where $V$ is the value function $V(x_1) = (1 - 2x_1)\, ln \frac{1-x_1}{x_1}$.

Based on the conservation of isotope mass and elemental mass, $M_P$ and $M_T$ can be obtained as

$$M_P = \frac{x_{1,F} - x_{1,T}}{x_{1,P} - x_{1,T}} M_F \quad (5)$$

$$M_T = \frac{x_{1,P} - x_{1,F}}{x_{1,P} - x_{1,T}} M_F \quad (6)$$

To obtain $x_{1,P}$ and $x_{1,T}$, we need to consider how $\vec{x}$ varies in the enriching process. The separation capability is evaluated by the separation factor $\alpha_0$, which equals to $\frac{x_{i,P}(n)}{1-x_{1,P}(n)} / \frac{x_{j,P}(n)}{1-x_{j,P}(n)}$, where the molar mass difference between the i th isotope and the j th isotope is 1 g/mol. Assuming that the separator is symmetric between product and tail, the isotope abundance of feed in stage $n$ and stage $n+1$ satisfies

$$R_{i,F}(n+1) = \frac{x_{i,F}(n+1)}{x_{1,F}(n+1)} = \frac{x_{i,F}(n)}{x_{1,F}(n)} \beta_{i,F} = R_{i,F}(n) \alpha_0^{-\frac{(m_i - m_1)}{2}} \quad (7)$$

Where $x_{i,F}(n)$ means the abundance of the $i$ th isotope at the feed of the $n$ th stage. $R_{i,F}(n)= \frac{x_{i,F}(n)}{x_{1,F}(n)}$, which represents the molar ratio between the $i$ th isotope and the lightest isotope at the feed of the $n$ th stage. $\beta_{i,F}$ is defined as the ratio of $R$ of the $n+1$ to stage to that of the $n$ th stage. Since $1 = \sum_i x_{i,F}(n+1) = x_{1,F}(n+1) \sum_i R_{i,F}(n+1)$, then $x_{i,F}(n+1)$, the abundance of the $i$ th isotope after stage $n+1$ can be expressed as

$$x_{i,F}(n+1) = x_{1,F}(n+1) R_{i,F}(n+1) = \frac{R_{i,F}(n+1)}{\sum_i R_{i,F}(n+1)} \quad (8)$$

By combining eq. (3) to (7), we will be able to obtain *SWU(n)* as a function of $\alpha_0$. $\alpha_0$ depends on the separation method and specific elements, and we will discuss it together with the cost per SWU ($C_{SWU}$) in the next paragraph.

$C_{SWU}$ is not easy to obtain, as little data is available and the production scale of many isotopes are << 1 ton/yr, so the current price cannot reflect that in large scale production. Interestingly, the two isotopes with the largest production scales are those with extreme masses: $^{2}$H (D) (>200 tons/yr)[21] and $^{235}$U (~260 tons/yr)[11, 12]. Their costs, which are \$1.33/SWU for D and \$38/SWU for $^{235}$U, are used to estimate the cost of other isotopes in large-scale production[10, 22, 23] (see S.I. Section S2). Moreover, D and $^{235}$U also represent two mainstream methods for large-scale isotope enrichment, chemical exchange for D and gas centrifugation for $^{235}$U. Hence, $\alpha_0$ is set as 2 for chemical exchange, corresponding to the separation factor for H/D separation[24], while $\alpha_0$ is taken as 1.5 for centrifugation, based on value reported for $^{235}$U enrichment[12].

To translate the cost of D and $^{235}$U to other elements, we also need to understand how cost scales with atomic weight and separation factor in each method, since they vary from element to element, and all these parameters influence $C_{SWU}$. In gas centrifugation, to simplify the model, we assume that the size, flow rate, temperature, and rotation rate are all the same as a centrifuge for uranium separation, so the operational cost is the same as uranium separation. Moreover, $\alpha_0$ becomes a constant for all elements under such assumptions. Moreover, $ln(\alpha) \propto \Delta m(n)$, and the production rate $\propto \bar{m}(n)$, where $\Delta m(n) = \frac{\sum m_i x_i(n)(i=2-k)}{1-x_1(n)} - m_1$, the mass difference between the lightest isotope and the abundance-weighted average of all other

isotopes in the *n* th stage, and $\bar{m}(n)$ is the average molar mass of all isotopes in the *n* th stage. Therefore, SWU $\propto (ln(\alpha))^2 \times \bar{m} \propto (\Delta m)^2 \times \bar{m}$, and $C_{SWU} \propto (\Delta m)^{-2} \times \bar{m}^{-1}$ (See S.I. section S3). Hence, $C_{SWU}$ for a given element *E* satisfies

$$C_{SWU}(E)/C_{SWU}(^{235}U) = \left(\frac{\Delta m_U}{\Delta m_E}\right)^2 \times \left(\frac{\bar{m}_U}{\bar{m}_E}\right) \quad (9)$$

In chemical separation, we also assume that the operational cost of a separator and the molar production rate are the same across different elements. Since $ln(\alpha)$ scales with $\Delta m/(m_1 \times (m_1 + \Delta m))$, and the production rate (kg/s) scales with $\bar{m}$, so SWU is proportional to $(ln(\alpha))^2 \times \bar{m}$ (See S.I. section S3) and $C_{SWU}(E)$ for a given element *E* satisfies

$$C_{SWU}(E)/C_{SWU}(D) = \frac{(\frac{1}{2})^2}{(\frac{\Delta m_E}{m_{1,E}(m_{1,E}+\Delta m_E)})^2} \times \left(\frac{\bar{m}_D}{\bar{m}_E}\right) \quad (10)$$

where ½ is the $\frac{\Delta m}{m_1(m_1+\Delta m)}$ value for D.

Then by combining eq. (3) to (10), we will be able to calculate $C(\vec{x})$ for enriching an element from natural abundance to the target isotopic composition $\vec{x}$. To evaluate whether the estimated cost is reasonable, we compared the cost of $^6$Li predicted by the model and the market price. Commercial price of 95% $^6$Li sold at 50 grams is ~\$58/g (See S.I. section S4), whereas our model predicts a cost of ~\$3.5/g. Such a ~10 X difference is reasonable since retail prices of chemicals are often 10-100 times that at large scale. For instance, the price of LiCl is approximately \$436/kg when purchased at 2.5 kg level, whereas industrial-scale acquisition at the ton scale reduces the cost to approximately \$12/kg (See S.I. section S4). Therefore, we think that our model captures the correct magnitude of enrichment cost at large scale and provides a reasonable basis for comparing potential elements.

## 3. Results

We analyzed 12 common elements that are used in aircraft which span over the entire periodic table, including Li, B, C, Mg, Cl, Ti, Ni, Fe, Cu, Zn, Mo, and Sn. We assume that Cl, Ti, Ni, Fe, Cu, Zn, Mo and Sn are enriched by gas centrifugation, and Li, B, C, and Mg are enriched by chemical exchange. This aligns with literature since $\alpha$ in chemical exchange scales with $\Delta m/(m_1 \times (m_1+\Delta m))$ while $\alpha$ in gas centrifuge scales with $\Delta m$ only[20], so the effectiveness of chemical exchange decreases sharply with increasing atomic mass.

### 3.1 Results without stripping stages.

Once an element and the separation method are determined, we combine eq. (2) to (10) to calculate $r(\vec{x})$, $C(\vec{x})$ and $G(\vec{x})$, respectively. First, we don't include any stripping stages in the enrichment cascade. Take Li and Mo as two examples, the results show that $G(1-r(\vec{x}))$ increases approximately linearly with $x_1$, since higher $x_1$ results in a larger saving in mass and thus the cost (Fig. 2). On the other side, $C(\vec{x})$ also increases with $x_1$ as more effort is needed to further enrich the lightest isotope. The $C(\vec{x})$ vs. $x_1$ curve is convex because at a higher $x_1$, it is more difficult to further increase $x_1$ for the same relative percentage of change compared to a smaller $x_1$.

Fig. 2 shows that the 12 target elements can be divided into three groups based on the maximum economic benefit $B(\vec{x})$. The first group is with a large economic benefits, such as Mo (\$175.7/kg at $x_{1,P}$ = 0.467, Fig. 2a), Sn (\$79.8/kg at $x_{1,P}$ = 0.046, Fig. 2b), Li (\$73.8/kg at $x_{1,P}$ = 0.180, Fig. 2c), Zn (\$67.6/kg at $x_{1,P}$ = 0.660, Fig. 2d), Ni (\$26.8/kg at $x_{1,P}$ = 0.758, Fig.

2e), and B (\$25.7/kg at $x_{1,P}$ = 0.245, Fig. 2f). The underlying reason is that these elements have a large $\Delta m$ if separated by gas centrifugation or a large $\Delta m/(m_1 \times (m_1+\Delta m))$ if separated by chemical exchange. Moreover, the relative portion of the lightest element is typically moderate (e.g., between 10% and 70%), so it is easy to increase $x_1$ by separation. The exception is Sn with $x_1$ = 0.97%. The reason is that $^{116}$Sn, a relatively light isotope of Sn, is 14.5%, and the mass difference between $^{116}$Sn and the average of other heavier isotopes is still as large as 2.88, ensuring efficient separation.

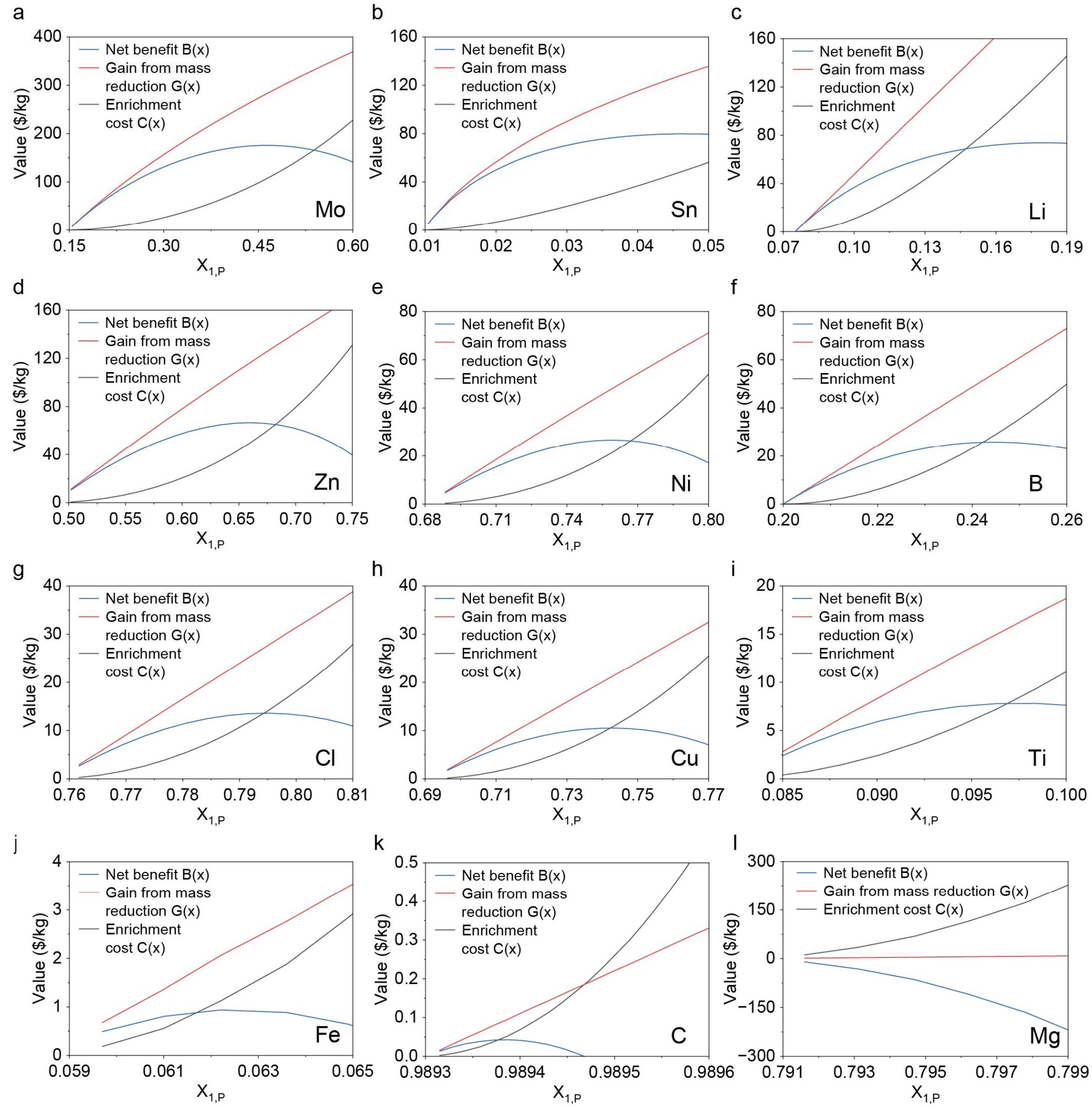


Figure 2. *B(x)*, *G(x)* and *C(x)* as a function of $x_{1,P}$ for all 12 elements studied: a) Mo; b) Sn; c) Li; d) Zn; e) Ni; f) B; g) Cl; h) Cu; i)Ti; j) Fe; k) C; l) Mg. *B(x)* is the net economic benefit of replacing 1 kg of an element with light-isotope-enriched one (blue line). *G(x)* is the economic gain from saved weight *G* (red line). *C(x)* is the cost of enrichment *C* (black line).

The second group includes Cl ($13.6/kg at $x_{1,P}$ = 0.795, Fig. 2g), Cu ($10.5/kg at $x_{1,P}$ =

0.744, Fig. 2h), and Ti (\$7.8/kg at $x_{1,P}$ = 0.098, Fig. 2i), and where the economic gain is moderate. This is because the lightest isotope's abundance is close to either 100% or 0%, such as 75% for $^{35}$Cl and 8% for $^{46}$Ti. Moreover, Δm for Cl and Ti are 2 and 2.1, respectively, which is not efficient for gas-centrifuge-based separation. Therefore, it is not effective to increase $x_1$ for these isotopes by enrichment. The last groups include Fe (\$0.9/kg at $x_{1,P}$ = 0.062, Fig. 2j), C (\$0.04/kg at $x_{1,P}$ = 0.9894, Fig. 2k), and Mg (\$0/kg, Fig. 2l). The reason is that $^{12}$C, the lightest isotope of C, is already 98.93% in abundance, making further enrichment not effective for weight reduction. The low benefit of Fe comes from the low natural abundance of the lightest isotope $^{54}$Fe ($x_{1,nat}$ = 0.059) and a low $\Delta m$ of 2, which lead to a much higher SWU to achieve the same amount of reduced mass. The non profitability of Mg is due to the extremely low $\frac{\Delta m}{m_1(m_1+\Delta m)}$, which increases the cost even at a low enrichment level.

### 3.2 Economic analysis with stripping stages

The analysis above is based on no stripping stage, which maximizes the profit since no extra cost is needed for stripping stages. However, this leads to a large amount of feed materials, since a substantial amount of the lightest isotope remains in the depleted end. This point is reflected in eq. (4), as $x_{1,F}-x_{1,T}$ is low and thus $M_T/M_P$ is ultrahigh. This issue can be addressed by adding stripping stages in the cascade and reduce $x_{1,T}$, the abundance of the lightest isotope in the tail. The dependence of $B(\vec{x})$ and the amount of feed materials $M_F$ on the tail-to-feed ratio $x_{1,T}/x_{1,F}$ are shown in Fig. 3. In each plot, the optimal $x_{1,P}$ for each element in Fig. 2 are used. It is clear that when $x_{1,T}/x_{1,F}$ decreases, $B(\vec{x})$ only decreases moderately but $M_F$ decreases

sharply. By taking the balance of economic benefits and the consumption of feed materials into account (e.g., $M_F$ starts to decrease slowly), table 1 gives the new recommended $x_{1,T}/x_{1,F}$ for each element with corresponding benefit $B(\vec{x})$ and materials consumption, which are also marked as red dash lines in Fig. 3.

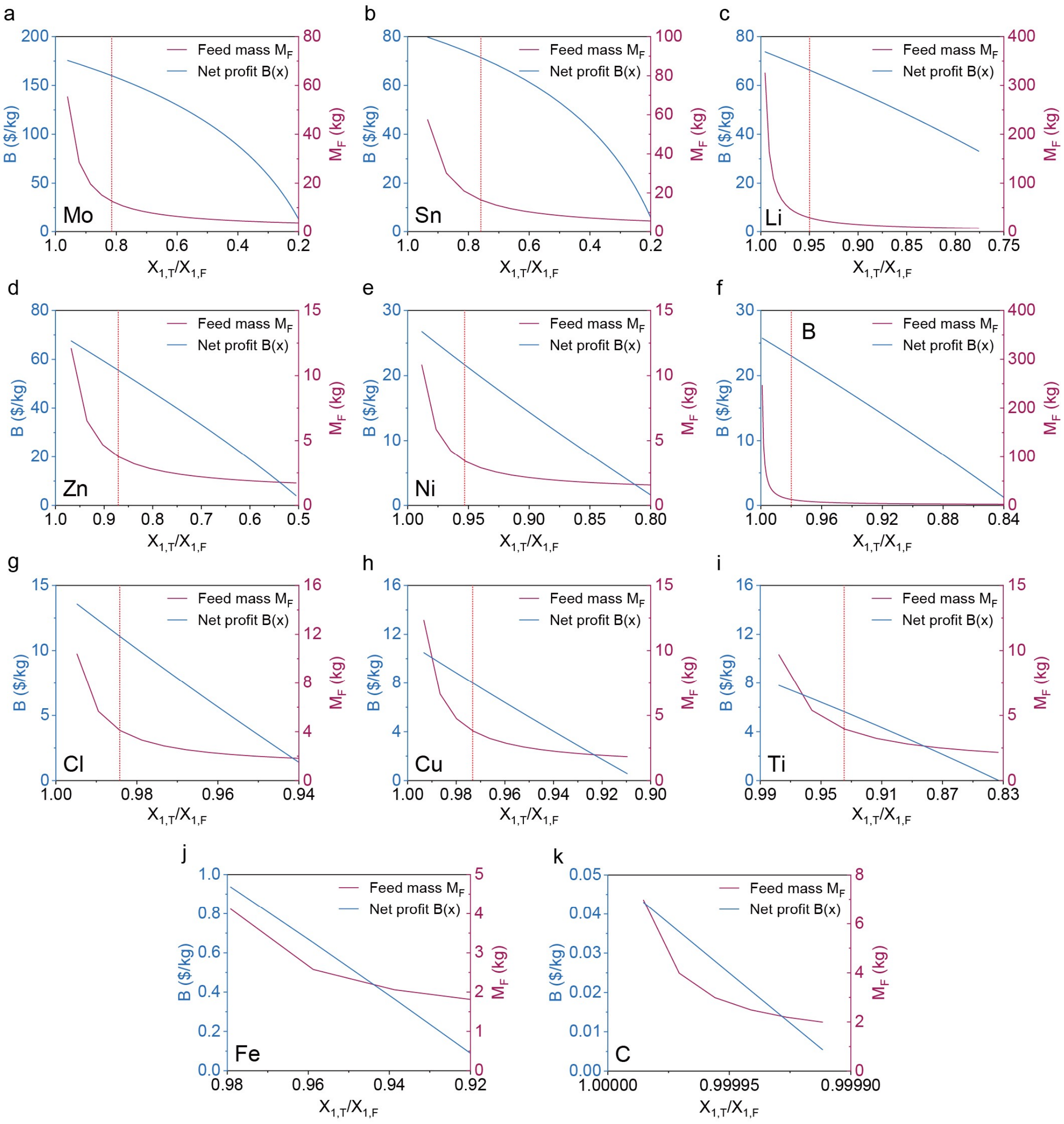


Figure 3. The feed mass $M_F$ and the corresponding net economic benefit of replacing 1kg of an element with light-isotope-enriched one $B(x)$ as a function of $x_{1,T}/x_{1,F}$. a) Mo, b) Sn, c) Li, d)

Zn, e) Ni, f) B, g) Cl, h) Cu, i) Ti, j) Fe, and k) C. The red dash line indicated the recommended $x_{1,T}/x_{1,F}$ for each element and thereby the corresponding benefit $B(x)$.

Table 1. Optimal net economic benefit $B$ with and without feed mass $M_F$ taken into account.

| Element | $x_{1,nat}$ | $x_{1,P}$ | No stripping stages | | With stripping stages | | |
|---|---|---|---|---|---|---|---|
| | | | $M_F/M_P$ | $B$ ($/kg) | $x_{1,T}/x_{1,F}$ | $M_F/M_P$ | $B$ ($/kg) |
| C | 98.93% | 98.94% | 7.0 | 0.04 | N/A | 7.0 | 0.04 |
| Li | 7.53% | 18.03% | 326 | 74 | 0.949 | 29 | 66 |
| B | 25.26% | 24.48% | 247 | 26 | 0.980 | 12.3 | 23 |
| Cl | 76.17% | 79.80% | 10.4 | 13.6 | 0.984 | 4.1 | 11.1 |
| Ti | 8.44% | 9.84% | 9.7 | 7.8 | 0.935 | 3.9 | 5.6 |
| Fe | 5.97% | 6.22% | 4.1 | 0.94 | N/A | 4.1 | 0.94 |
| Ni | 68.85% | 75.84% | 10.8 | 27 | 0.952 | 3.4 | 22 |
| Cu | 69.63% | 74.38% | 12.3 | 10.5 | 0.973 | 3.8 | 8.8 |
| Zn | 50.19% | 66.02% | 12.1 | 68 | 0.870 | 3.8 | 55 |
| Mo | 15.44% | 46.72% | 55 | 176 | 0.813 | 12.5 | 160 |
| Sn | 1.04% | 4.57% | 57 | 80 | 0.760 | 16.5 | 71 |

With the optimal level of enrichment described in Table 1, now we can estimate the cost saving for an airplane and other transport vehicles. For example, an Airbus A380 airplane contains approximately 95 kg of Li, 410 kg of Fe, 3,195 kg of Cu, and 6,566 kg of Zn in

different Al alloys used to construct the airframe structure[6, 25-28]. In addition, 145 kg of Ti, 6,820 kg of Ni, and 409 kg of Mo are used in the Ni superalloy for the turbofan engines[29-31], and 27,600 kg of Ti is employed in corrosion critical parts and joints within composite assemblies[7]. Replacing the current materials by the optimal isotopic lightened material in Table 1 can reduce material usage of 1.4 kg of Li, 36 kg of Ti, 0.06 kg of Fe, 24.6 kg of Ni, 5.2 kg of Cu, 58 kg of Zn, and 8.8 kg of Mo, resulting in an estimated mass saving of 134.2 kg (Fig. 4a) and cost savings of ~$764 K (Fig. 4d) for a single A380 aircraft over a 30-year service lifetime (See S.I. section S5). Similarly, significant mass reduction can also be achieved in spacecraft applications. For instance, SpaceX Falcon 9 can achieve a mass reduction of 19.6 kg (Fig. 4b) and profit of over $516 K (Fig. 4e) by replacing elements such as Li, Cu, Ni, and Mo. SpaceX Starship can reduce the mass by 102.5 kg (Fig. 4c) and generate profit of $2.37 million (Fig. 4f) by replacing Fe, Ti, Ni, and Mo (see S.I. section S1, S6, and S7).

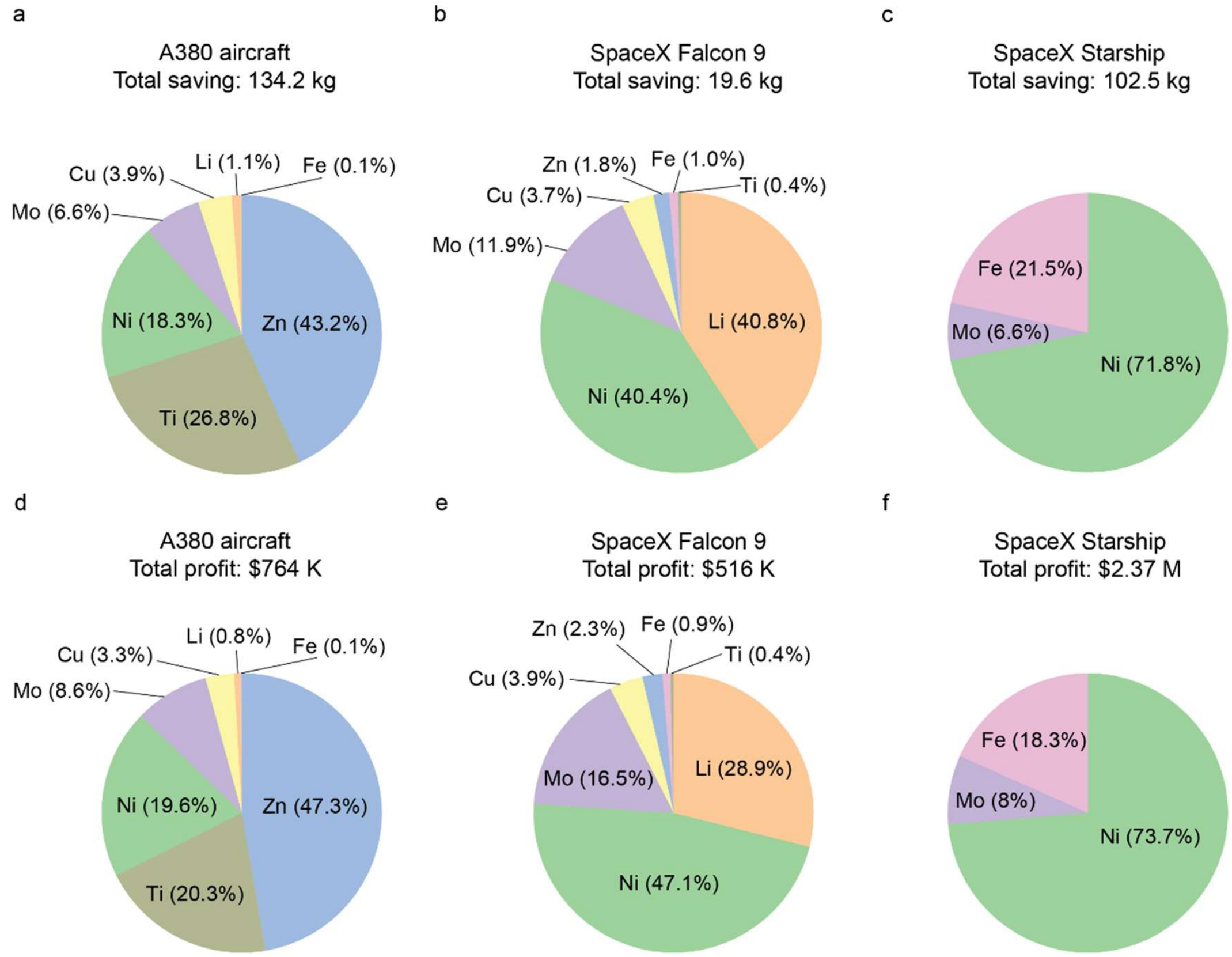


Figure 4. Elemental contribution to total mass saving and economic profit for three representative vehicles: A380 aircraft (a, d), SpaceX Falcon 9 (b, e), and SpaceX Starship (c, f). Panels a to c show mass saving contributions, while panels d to f show profit contributions. Total mass saving and total profit are labeled above.

Currently, Zn, Ni and Mo have been produced by gas centrifugation for nuclear fission reaction and medical isotope precursors[15-17]. The target isotopes are the heavier ones inside (e.g., $^{64}Zn$ depleted, $^{64}Ni$, $^{98}Mo$/$^{100}Mo$). Therefore, the light ones enriched in the stripping stages can be considered as sources for such lightweight elements to increase value of the existing isotope enrichment process.

We also want to emphasize that this analysis is preliminary and semi-qualitative, and it is intended to provide an order-of-magnitude assessment of the potential benefit of isotopic mass reduction rather than a precise economic prediction. The enrichment cost $C(\vec{x})$ in this study is based on simplified scaling relationships derived from large-scale produced D and $^{235}U$, while the true cost depends strongly on the specific chemical precursors and process design. For example, different precursor compounds can lead to large differences in separation efficiency, operating temperature, and energy consumption, while process design can affect capital and maintenance cost. In addition, several elements considered in this work are not even currently produced at the kilogram scale or lack a clear pathway for scalable isotope enrichment (e.g., Cl, Sn)[32, 33], which further increases uncertainty. Therefore, the results should be interpreted as a screening tool that highlights the promising candidates, rather than a precise estimate of real-world enrichment costs.

**Acknowledgement**

This work was supported in part by the isotope program of the Department of Energy in the United States under Grant Number DE-SC0022256. We also acknowledge support from the internal grant in the School of Engineering and Applied Science at Columbia University.

**Supporting Information**

# Techno-economic Analysis of Light Isotope-enriched Elements for Lightweighting Applications

Wenbo Bao[1], Zhihao Yang[1], Taeyoung Wang[1], Joseph F. Wild[1], Yuan Yang[1]*

[1]Department of Applied Physics and Applied Mathematics, Columbia University, New York, NY 10027, USA. *Corresponding author. Email: yy2664@columbia.edu (Yuan Yang).

**Section S1. Derivation of Gain of Saving Fuels from Mass Reduction**

For aircraft:

Assuming two flights per day, the Federal Aviation Administration has reported that for an aircraft operating 20.4 hours per day, each additional pound added in aircraft weight leads to an increase of approximately 0.0061 gallons per hour in jet fuel consumption[1], which worths \$235 per kg weight saved.

$$0.0061\frac{gallons}{h*pound}\times 2.2\frac{pound}{kg}\times 20.4\frac{h}{day}\times 365\frac{day}{yr}\times \$2.35\frac{1}{gallon}=\$234.82/kg$$

The price of jet fuel has increased at an average annual rate of around 4.02% over the past 30 years[2]. Assuming an aircraft service life of 30 years, every 1 kg reduction in aircraft weight would lead to an estimated lifecycle fuel cost saving of \$13,212.

$$\$234.82/kg\times\frac{((4.02\%+1)^{30}-1}{4.02\%}=\$13{,}212/kg$$

For SpaceX Falcon 9 and Starship:

Although the current payload transportation cost of SpaceX Falcon 9 and Starship is as high as \$7000/kg[3], it is reasonable to assume that this cost could decrease in the future after larger-scale production, lower target launch costs, and full reusability[4]. Because Falcon 9 is only partially reusable, with its second stage expended after launch, its future transportation cost is assumed to remain higher, at approximately \$300/kg, with a lifetime reuse of 100 cycles[5]. In contrast, Starship is designed to be fully reusable, allowing both stages to be recovered and reused, so it is assumed to achieve a lower future transportation cost of \$100/kg[6]. Starship is

also assumed to support more reuse cycles, with a lifetime of 300 cycles, due to its design goal of aircraft-like reusability[7]. In this study, the economic benefit associated with saving 1 kg of SpaceX Falcon 9 and Starship therefore are both taken to be $30,000/kg.

## Section S2. Cost per SWU for Deuterium Separation and $^{235}U$ Separation

For chemical separation methods, the reference cost is based on the market price of 99.8% purity of industrial-scale produced deuterium oxide $D_2O$, which is approximately $2/g[8]. This corresponds to an estimated price of $10/g for 99.8% D. Based on the assumption that the tail concentration of D for H/D separation is 70% of the natural abundance, the concentration of D in the feed, tail, and product streams is 0.0156%, 0.011%, and 99.8%, respectively. According to eq. (4) to (6) in the main text, producing 1 kg of 99.8% D enriched H/D mixture requires around 7,593 SWU. Hence, the cost for D enrichment is $1.33/SWU*kg.

As for centrifuge method, the separation cost of $^{235}U$ per SWU is already available from commercial reports. In 2018, which represents a more reasonable reference year compared with recent year, the SWU price fluctuated between $36/SWU and $40/SWU[9, 10]. Accordingly, the mean value of $38/SWU for $^{235}U$ enrichment is used as the baseline estimate in this work.

**Section S3. Derivation of the Cost Factor for Centrifugation and Chemical Exchange**

In a single stage separation, based on the definition of the separation factor $\alpha$ and the assumption that $\alpha$ is close to 1, we have

$$\alpha = \frac{\frac{x_{1,P}}{1 - x_{1,P}}}{\frac{x_{1,T}}{1 - x_{1,T}}}$$

For centrifugation separation, based on two shell profile[11], we have:

$$\alpha = \exp\left(\frac{\Delta m(\omega^2(r_0^2 - r_1^2)^2}{2RT}\right)$$

$$\varepsilon = \alpha - 1 \approx ln\alpha \propto \Delta m$$

For chemical exchange, based on the chemical equilibrium[12], we have:

$$\varepsilon = \alpha - 1 \approx ln\alpha \propto -\Delta G \propto -\Delta E_0 \propto \frac{\Delta m}{m_1(m_1 + \Delta m)}$$

Where $\Delta E_0$ is the zero-point energy difference and $\Delta G$ is the Gibbs free energy change.

Then the enrichment process in one stage is as follows,

$$x_{1,P} = x_{1,F} + \delta x_{1,P}$$

$$x_{1,T} = x_{1,F} + \delta x_{1,T}$$

Let $g(x_1) = ln\frac{x_1}{1-x_1}$:

$$ln\alpha = g\left(x_{1,P}\right) - g\left(x_{1,T}\right)$$

$$g(x_{1,F} + \delta x_{1,P}) \approx g\left(x_{1,F}\right) + g'\left(x_{1,F}\right)\delta {x_{1,P}}^2$$

$$g(x_{1,T} + \delta x_{1,P}) \approx g\left(x_{1,F}\right) + g'\left(x_{1,F}\right)\delta {x_{1,T}}^2$$

$$ln\alpha \approx \frac{\delta x_{1,P} - \delta x_{1,T}}{x_{1,F}(1 - x_{1,F})}$$

$$\delta x_{1,P} \approx \frac{M_T}{M_F} x_{1,F}(1 - x_{1,F})\varepsilon$$

$$\delta x_{1,T} \approx -\frac{M_P}{M_F} x_{1,F}(1 - x_{1,F})\varepsilon$$

$$SWU = M_P V\left(x_{1,F} + \delta x_{1,P}\right) + M_T V\left(x_{1,F} + \delta x_{1,T}\right) - M_F V\left(x_{1,F}\right)$$

$$SWU = M_P \left[V\left(x_{1,F}\right) + V'\left(x_{1,F}\right)\delta x_{1,P} + \frac{1}{2} V''\left(x_{1,F}\right)\delta x_{1,P}{}^2\right]$$
$$+ M_T \left[V\left(x_{1,F}\right) + V'\left(x_{1,F}\right)\delta x_{1,T} + \frac{1}{2} V''\left(x_{1,F}\right)\delta x_{1,T}{}^2\right] - M_F V(x_{1,F})$$

$$SWU \approx \frac{1}{2} V''\left(x_{1,F}\right)\left[M_P \delta x_{1,P}{}^2 + M_T \delta x_{1,T}{}^2\right] \propto \varepsilon^2$$

Where $\delta x_{1,P}$ and $\delta x_{1,T}$ are the concentration change in the product stream and tail stream, respectively. While the total cost is fixed, the cost per SWU is proportional to 1/SWU. Meanwhile, the production rate (kg/s) scales with $\bar{m}$. Hence, we have

$$C_{SWU}(E)/C_{SWU}(^{235}U) = \left(\frac{\Delta m_U}{\Delta m_E}\right)^2 \times \left(\frac{\bar{m}_U}{\bar{m}_E}\right).$$

For centrifugation using $^{235}$U as the reference.

$$C_{SWU}(E)/C_{SWU}(D) = \frac{(\frac{1}{2})^2}{(\frac{\Delta m_E}{m_1(m_1 + \Delta m_E)})^2} \times \left(\frac{\bar{m}_D}{\bar{m}_E}\right)$$

For chemical exchange using D as the reference.

**Section S4. Comparison of Price between Lab-scale and Industrial-scale Produced Chemicals**

Table S1. Chemical market price at small scale and large scale.

| Chemical | Small-scale Price ($/kg) | Large-scale price ($/kg) |
|---|---|---|
| $^{6}Li$ | 58 | 3.5 (estimated by this work) |
| LiCl | 436 | 12 |
| $TiCl_4$ | 20 | 0.5 |
| $SnCl_2 \cdot 2H_2O$ | 520 | 10 |

Links to price sources:

$^{6}Li$:

Small scale (50 g):

https://www.chemicalbook.com/Price/LITHIUM-6.htm

LiCl:

Small scale (2.5 kg):

https://www.sigmaaldrich.com/US/en/product/sigald/793620?srsltid=AfmBOoqRBUjEIT7FjSGcjGMujRH5p4zxBGEh4Z85U1PzDq-9vmLB2WP3

Large scale (1 ton):

https://www.alibaba.com/product-detail/Fast-Delivery-Best-Price-LiCl-Anhydrous_1600552057052.html?spm=a2700.details.you_may_like.2.6fb48621R4yi85

$TiCl_4$:

Small scale (1.5 kg):

https://www.sigmaaldrich.com/US/en/product/sigald/208566?srsltid=AfmBOoo-xY9NlOPsSWQoFfUGXhifAJfE12fUY8vuWywSR5V469Zh6KBU

Large scale (15 tons):

https://www.alibaba.com/product-detail/Industrial-Grade-Ticl4-Price-99-9_11000014217839.html?spm=a2700.7724857.0.0.420933d5VT1jFp

$SnCl_2 \cdot 2H_2O$:

Small scale (2.5 kg):

https://www.sigmaaldrich.com/US/en/product/sigald/243523?srsltid=AfmBOorNPYC9XJfgCRveLRKdfCubZDOWridVbk7lCb02Wf0SCKKv7w2y

Large scale (1 ton):

https://www.alibaba.com/product-detail/99-min-SnCl2-2H2O-Stannous-Chloride_60717482874.html?spm=a2700.galleryofferlist.normal_offer.d_title.66d113a0evceNU&priceId=5aff731c04394974ab24da2ea67fa0d2

**Section S5. Estimation of Cost Savings for an A380 Airplane**

The operating empty weight (OEW) of an A380 airplane is 276,000 kg[13]. It is reported that the Ti constitutes approximately 10% of the total mass of the A380, corresponding to 27,600 kg[14]. Furthermore, various Al alloys used in different sections of the A380 account for 61% of the total mass[15]. Although the detailed distribution of specific alloy has not been disclosed, a reasonable assumption is that roughly 75% of Al alloys consists of 7xxx series alloy used in primary airframe structure, along with Al-Li alloys in selected weight-critical components, while the remaining 20% consisting of 2xxx and 6xxx series alloys used in other structural or functional components[16, 17]. The estimated proportions of isotopic weight-reducible materials for different Al alloys are summarized in Table 2.

Table S2. Composition of different Al alloys and total mass of materials used in Al alloys on an A380 airplane.

| Material | Alloy | | | Total Mass (kg) |
|---|---|---|---|---|
| | 7xxx[17] | 2xxx/6xxx[17] | Al-Li[18] | |
| Li | - | - | 1.5% | 95 |
| Fe | 0.3% | 0.5% | - | 410 |
| Cu | 2.0% | 4.4% | 3.0% | 3,195 |
| Zn | 6.9% | - | 0.5% | 6,566 |

A typical A380 engine such as the Rolls-Royce Trent 900 has a dry mass of about 6,200 kg, and it is indicated that around 50% of aircraft turbine engines consist of Ni-based superalloys[19, 20]. Using representative alloy compositions (~55wt% of Ni, 3.3wt% of Mo, and 1.15wt% of Ti)[21], the four engines together correspond to roughly 6,820 kg of Ni, 409 kg of Mo, and 145kg Ti. Overall, replacing the material mentioned above can reduce the material usage of 1.4 kg of Li, 36 kg of Ti, 0.06 kg of Fe, 24.6 kg of Ni, 5.2 kg of Cu, 58 kg of Zn, and 8.8 kg of Mo, resulting in a cost saving of $764,200 per A380 airplane.

## Section S6. Estimation of Cost Savings for SpaceX Spacecrafts

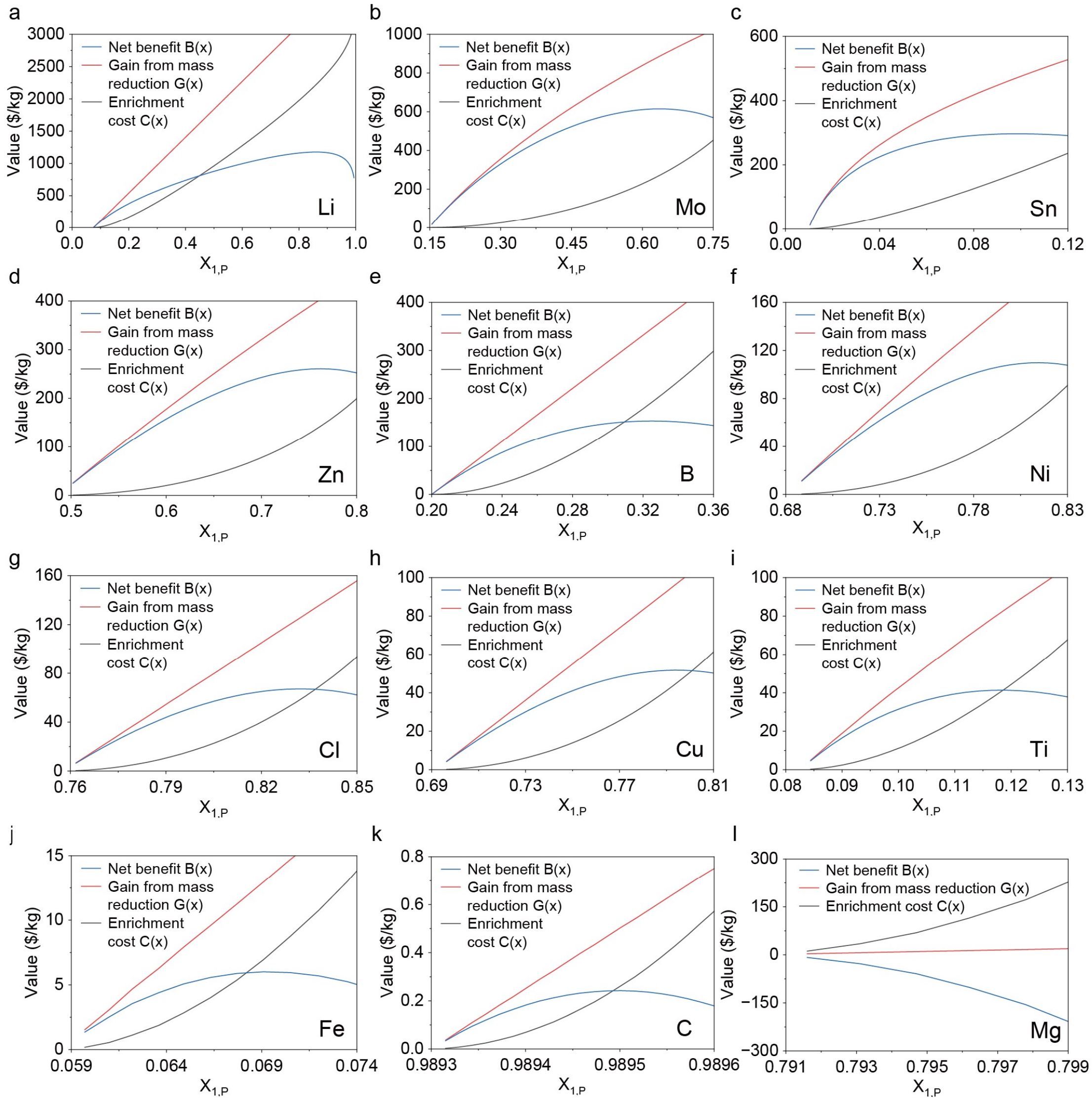

Figure S1. *B(x)*, *G(x)* and *C(x)* as a function of $x_{1,P}$ for all 12 elements studied for SpaceX spacecrafts: a) Mo; b) Sn; c) Li; d) Zn; e) Ni; f) B; g) Cl; h) Cu; i) Ti; j) Fe; k) C; l) Mg. *B(x)* is the net economic benefit of replacing 1 kg of an element with light-isotope-enriched one (blue line). *G(x)* is the economic gain from saved weight *G* (red line). *C(x)* is the cost of enrichment *C* (black line).

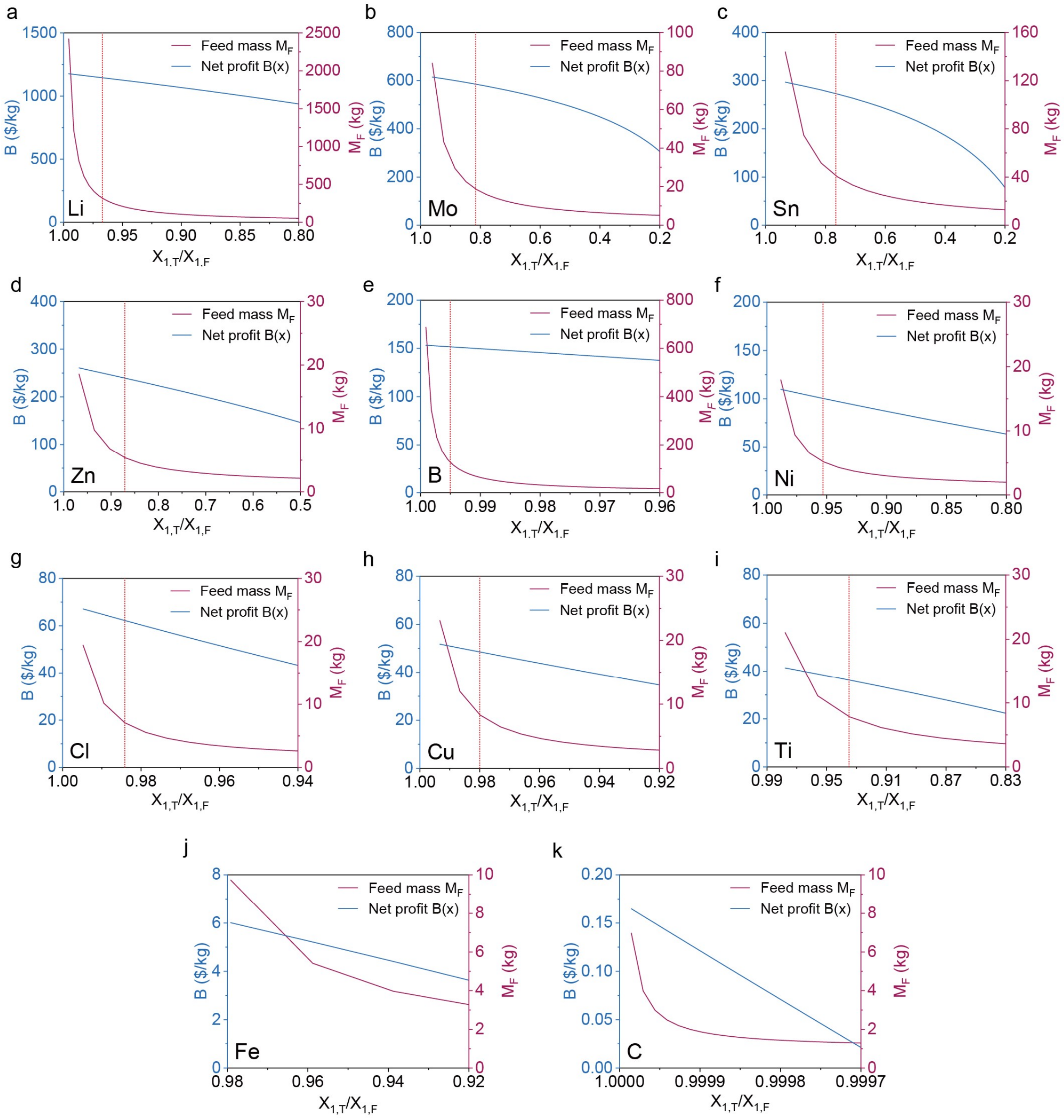

Figure S2. The feed mass $M_F$ and the corresponding net economic benefit of replacing 1kg of an element with light-isotope-enriched one $B(x)$ as a function of $x_{1,T}/x_{1,F}$ for SpaceX spacecrafts. a) Mo, b) Sn, c) Li, d) Zn, e) Ni, f) B, g) Cl, h) Cu, i) Ti, j) Fe, and k) C. The red dash line indicated the recommended $x_{1,T}/x_{1,F}$ for each element and thereby the corresponding benefit $B(x)$.

Table S3. Optimal net economic benefit $B$ with and without feed mass $M_F$ taken into account for SpaceX spacecrafts.

| Element | $x_{1,nat}$ | $x_{1,P}$ | No stripping stages | | With stripping stages | | |
|---|---|---|---|---|---|---|---|
| | | | $M_F/M_P$ | $B$ ($/kg) | $x_{1,T}/x_{1,F}$ | $M_F/M_P$ | $B$ ($/kg) |
| C | 98.93% | 98.94% | 7.0 | 0.16 | N/A | 7.0 | 0.16 |
| Li | 7.53% | 86.04% | 2421.2 | 1176.4 | 0.970 | 350.9 | 1148.6 |
| B | 25.26% | 32.50% | 687.9 | 153.3 | 0.995 | 138.6 | 151.8 |
| Cl | 76.17% | 83.08% | 19.4 | 67.1 | 0.984 | 7.1 | 62.2 |
| Ti | 8.44% | 11.92% | 21.0 | 41.3 | 0.935 | 7.8 | 36.1 |
| Fe | 5.97% | 6.91% | 9.7 | 6.0 | N/A | 9.7 | 6.0 |
| Ni | 68.85% | 81.48% | 18.0 | 109.9 | 0.952 | 5.1 | 100.3 |
| Cu | 69.63% | 79.33% | 23.1 | 51.8 | 0.973 | 6.5 | 46.9 |
| Zn | 50.19% | 76.23% | 18.6 | 260.8 | 0.870 | 5.4 | 239.3 |
| Mo | 15.44% | 63.57% | 84.1 | 615.1 | 0.813 | 18.6 | 584.9 |
| Sn | 1.04% | 10.08% | 143.9 | 296.8 | 0.760 | 40.1 | 272.2 |

**Section S7. Estimation of Cost Savings for SpaceX Falcon 9 and Starship**

We assume the shipping price of Falcon 9 is 300/kg and one Falcon 9 spacecraft can perform 100 trips in 10 years of service, so saving 1 kg of mass corresponds to an extra revenue of $30,000 during its whole life, as discussed in section S1.

The dry mass of SpaceX Falcon 9 is 29,500 kg, of which around 25,600 kg corresponds to the recoverable portion (e.g., first-stage booster and payload fairing, fuel not included) [22, 23]. Within this portion, approximately 13,000 kg of Al-Li AA 2198 alloy is used in primary structural framework[24], containing 1.0 % of Li, 3.2 % of Cu, and 0.38% of Zn[18]. Using optimized enriched light isotope in Table S3 could save 14.7 kg Li, 1.3 kg Cu, and 0.66 kg Zn, which generate a profit of $181 K. In addition, nine Merlin 1D engines can be recovered[22]. Together, these engines contain approximately 4,410 kg of IN718 alloy[20, 25], which consists of 3.3% of Mo, 1.15% of Ti, 18.5% of Fe, and 55% of Ni[21]. Using optimized enriched light isotope in Table S3 could save 0.14 kg Ti, 0.35 kg Fe, 4.3 kg of Mo and 14.6 kg Ni, which generate a profit of $335 K over its whole life. So, the total saving is $516 K by using isotope-enriched lightweight materials.

Similarly, we assume the shipping price of the future Starship is 100/kg and one Starship spacecraft can perform 300 trips in 15-20 years of service, so saving 1 kg of mass corresponds to an extra revenue of $30,000 during its whole life, as discussed in section S1.

The fully reusable SpaceX Starship has a dry mass of 120 t and is equipped with six Raptor engines[26]. Based on recent test configurations and the expected launch timeline, the Raptor 2 engine with a mass of 1,630 kg is likely to be similar to the final engine configuration[27-29]. Therefore, the IN718 alloy used in these engines could reduce 0.31 kg Ti, 0.78 kg Fe, 9.5 kg of Mo and 32.3 kg Ni, and generate a profit of $743 K. In addition, because the main body of Starship is entirely made of stainless steel, after excluding the mass of the engines and other components, it is reasigureonable to estimate that approximately 100 t of stainless steel is used in Starship[30]. The most suitable 304L stainless steel contains 70.5% Fe and 12% Ni[31], which could save 30.5 kg of Fe and 72.0 kg of Ni, which generate a profit of $1.63 million. Therefore, after a lifetime of 100 reuse cycles, the use of isotope-enriched lightweight materials could save $2.37 million for Starship spacecrafts.